\documentclass[aps,prd,eqsecnum,11pt,showpacs,preprintnumbers,nofootinbib,superscriptaddress,longbibliography]{revtex4-2}
\usepackage{geometry}                
\usepackage[pdftex]{graphicx}
\usepackage{float}
\usepackage{amssymb,amsmath,amsopn,bm,dsfont}
\DeclareMathAlphabet{\matheul}{U}{eus}{m}{n}
\usepackage{setspace,enumitem}

\allowdisplaybreaks

\usepackage[colorlinks,pdfstartview=FitH]{hyperref}
\hypersetup{linkcolor=blue,citecolor=blue,filecolor=black,urlcolor=blue}

\newcommand{\EE}{{\matheul E}}
\newcommand{\PP}{{\matheul P}}

\begin{document}

\title{Rotating anisotropic stringy spheroid in a modified Hartle formalism} 
\author{Philip Beltracchi}
\email{philipbeltracchi@gmail.com}
\affiliation{Department of Physics and Astronomy, University of Utah\\Salt Lake City, Utah 84112, USA}

\begin{abstract}
\noindent Here we look at an application of the Hartle metric to describe a rotating version of the spherical string cloud/ global monopole solution. While rotating versions of this solution have previously been constructed via the Newman-Janis algorithm, that process does not preserve the equation of state. The Hartle method allows for preservation of equation of state, at least in the sense of a slowly rotating perturbative solution. In addition to the direct utility of generating equations which could be used to model a region of a rotating string cloud or similar system, this work shows that it is possible to adapt the Hartle metric to slowly rotating anisotropic systems with Segre type [(11)(1,1)] following an equation of state between the distinct eigenvalues.
\end{abstract}

\begin{description}
\item[PACS numbers]
\end{description}
\maketitle

\section{Introduction}
We use the shorthand ``koosh" to describe the hyperconical, spherically symmetric Kerr-Schild geometry with the line element
 \begin{equation}
     ds^2=-\kappa^2  dt^2+\frac{1}{\kappa^2}dr^2+ r^2 d\theta^2+r^2 \sin^2\theta \, d\phi^2.
   \label{uniaxialmet}
 \end{equation}
 This metric describes a hypercone in four dimensions in that the circumference of a circle of proper radius $r^*$ is $2\pi \kappa r^*$. Its Riemann tensor has only one independent nonzero element $R^{\theta \phi}_{~~\theta \phi}=(1-\kappa^2)/r^2$. 
Since spherically symmetric Kerr-Schild metrics may be written in the form $-g_{tt}=1/g_{rr}=1-2m(r)/r$, where $m(r)$ is a ``mass" function, we can identify that for a koosh
 \begin{equation}
    m=\lambda r,
    \label{uniaxialmass}
 \end{equation}
such that $\kappa=\sqrt{1-2  \lambda}$. 
 Demanding that the $t$ coordinate remains timelike requires 
 $ 2\lambda<1$.  
 
 The only nonzero energy-momentum tensor components are
 \begin{align}
 T^t_{~t}=T^r_{~r}=-\frac{\lambda}{4\pi r^2}
 \end{align}
 This means that the eigenvalue structure is Segre type [(11)(1,1)] with
 \begin{align}
     \Lambda_0=\Lambda_1,\quad \Lambda_2=\Lambda_3=0
     \label{stringeos}
 \end{align}
 where $\Lambda_0$ is the eigenvalue associated with a timelike eigenvector and $\Lambda_{1,2,3}$ are the other eigenvalues. This can be thought of as a ``stringy" equation of state, in that it applies to solutions with vacuum cosmic strings \cite{PhysRevD.23.852,PhysRevD.31.3288,Letelier:1979ej,gursesnambu}.
 
This solution seems to have been initially discovered by Lettelier as a cloud of radially aligned strings \cite{Letelier:1979ej}, arranged like the filaments on a koosh ball toy and giving rise to our name. It was independently examined as a model for a ``global monopole"\cite{PhysRevLett.63.341}. Several later papers have also examined the koosh or similar systems as stringy systems or monopoles arising from various theories \cite{Guendelman:1991qb,Delice:2003zp,2019arXiv191008166B}. 
One intriguing recent development is the consideration of the $\lambda\rightarrow1/2^-$ limit. If such a system is cut off at a finite radius by a shell, it acts as an interior to the Schwarzschild black hole, and further has the correct mass scaling characteristics without changing the interior density profile. This is a ``quasiblack hole" configuration \cite{Lemos:2020hsz}.  The quasiblack hole does have the problem that it is a singular configuration, but it is noteworthy that the hyperconical geometry of metric Eq.~\eqref{uniaxialmet} only applies to global monopoles at sufficient radius from the center; at extremely small radii the global monopole described in \cite{PhysRevLett.63.341} is de Sitter like and nonsingular. Replacing the interior of a quasiblack hole with a very extreme global monopole would lead to a system with some properties like certain gravastar \cite{MazurMottola:2001,MazurMottola:2004,MazurMottola:2015,Visser:2003ge} or dark energy star models \cite{ChapHohlLaughSant:2001} (in that the compact object is bounded by some kind of thin shell at or near the horizon) and other properties like Bardeen \cite{Bardeen:1968} or similar (e.g.\cite{Dymnikova:1992,AyonBeato:1998ub,Hayward:2005gi}) type nonsigular black holes (in that the interior is Kerr-Schild and the pressure everywhere follows $p_r=-\rho$, there is a de Sitter center, and the solution decreases in density as one moves outward).

There have also been various examinations of rotating solutions generated from the Newman-Janis algorithm \cite{Newman:1965tw,Newman:1965my,Gurses:1975vu} which have string cloud behavior in their static versions \cite{Sakti:2019iku}. While the Newman-Janis algorithm preserves Segre type [(11)(1,1)], the stringy equation of state is not preserved in passing to rotation under the Newman-Janis algorithm \cite{Beltracchi2021a}.

In this paper we modify the Hartle formalism \cite{Hartle:1967he,Hartle:1968si} to produce a perturbative model for a slowly rotating koosh which preserves the stringy equation of state.  Originally, the Hartle formalism involved perfect fluid (Segre type [(111),1]) equations of state. Effects from first order in rotation for anisotropic systems had been previously considered for anisotropic neutron star models \cite{2015CQGra..32n5008S} and anisotropic continuous pressure gravastar models \cite{2008PhRvD..78h4011C}. Very recently, a treatment more similar to Hartle's involving the second order in rotation deformation terms for a particular anisotropic Segre type [(11)1,1] neutron star model was presented \cite{2021EPJC...81..698P}. The situation with a koosh is extremely anisotropic in that one of the distinct eigenvalues is always zero, and the Segre type is different than what has been considered previously.

\section{Axisymmetric spacetimes and the Hartle Formalism }
\label{Sec:Eineqs}

One convenient notation of the general axisymmetric metric in coordinates $(t,r,\theta,\phi)$ comes from \cite{Chandrasekhar:1984}
\begin{align}
ds^2 = -e^{2\nu} dt^2 + e^{2\psi} \big(d \phi - \omega dt\big)^2 + e^{2 \alpha} dr^2 + e^{2 \beta} d\theta^2
\label{axisymstat}
\end{align}
where the five functions $\nu, \psi, \alpha, \beta, \omega$ are functions of $r$ and $\theta$.  The functions $\nu,\psi,\omega$ can be isolated as scalar functions because of the existence of the time and axial Killing vectors 
\begin{align}
    K_{(t)}^\mu=(1,0,0,0),\\
    K_{(\phi)}^\mu=(0,0,0,1).
\end{align}
Adopting the nomenclature from \cite{2022PhRvD.105b4001B}, we define additional vectors $l$ and $N$, which in these coordinates are
\begin{align}
    l^\mu=(1,0,0,\omega),\\
    N_\mu=(-1,-e^{\alpha-\nu},0,0),
\end{align}
such that
\begin{align}
    l=K_{(t)}+\omega K_{(\phi)},\quad N\cdot N=0,~N\cdot l=-1.
\end{align}
With these auxiliary vectors, we can define two physically relevant scalar quantities, being the surface gravity parameter 
\begin{align}
    S_G=N_\mu l_\nu(\nabla^\mu l^\nu)=\frac{1}{2}e^{-\alpha-\nu}\frac{\partial}{\partial r}e^{2\nu}
\end{align}
and angular momentum density parameter
\begin{align}
    \mathcal{J}=-N_\mu l_\nu(\nabla^\mu K_{(\phi)}^\nu)=-\frac{1}{2}e^{2\psi-\alpha-\nu}\frac{\partial \omega}{\partial r}.
\end{align}
These scalars are related to the Komar mass and angular momentum(see \cite{Komar:1959} for the introduction of the concepts, and \cite{2022PhRvD.105b4001B} for information about the particular formulation), which can be defined as surface integrals at a given radius, or as the sum of a surface integral at a smaller radius and a volume integral of components of the energy-momentum tensor between the smaller and given radii
 \begin{align}
M_K(r) &=  \frac{1}{4 \pi G}\int_{\partial V_+}\!  (S_G + \omega \mathcal{J})\, dA
=\int_V \sqrt{-g}\  \Big(\! \!-T^t_{\ t} + T^r_{\ r}  + T^\theta_{\ \theta}  + T^\phi_{\ \phi}  \Big)\, dr\,d\theta\,d\phi\, +\, \frac{1}{4 \pi G} \int_{\partial V_-} \! (S_G + \omega \mathcal{J})\, dA, 
\label{KomarMass}\\
J_K(r) &=  \frac{1}{8 \pi } \int_{\partial V_+}\!  \! \mathcal{J} \, dA = \int_V  \sqrt{-g}\ T^t_{\ \phi} \,dr\,d\theta\,d\phi +\frac{1}{8 \pi } \int_{\partial V_-}\!  \!\mathcal{J} \, dA. 
\label{KomarAngmom}
\end{align}

For his perturbative framework, Hartle expanded the line element \eqref{axisymstat} to second order in the angular momentum as~\cite{Hartle:1967he}
\begin{align}
ds^2 = & - e^{2\nu_0(r)} \Big[ 1+2 h_0(r) + 2 h_2(r)\, P_2(\cos\theta)  \Big]  dt^2 \nonumber\\ & 
+ \frac{r}{r-2m(r)}  \left\{1+\frac{2}{r-2m(r)} \Big[ m_0(r) + m_2(r)\, P_2(\cos\theta) \Big] \right\}  dr^2 \nonumber\\& 
+ r^2 \Big[ 1+2 k_2(r)\, P_2(\cos\theta)  \Big] \Big[ d\theta^2 + \sin^2\!\theta\,\big(d\phi - \omega(r) dt \big)^2 \Big].
\label{kingmet}
\end{align}
The function $\omega(r)$ is the first-order contribution that gives rise to inertial frame dragging. Here $P_l(\cos\theta)$ is the Legendre polynomial of order $l$, $m(r)$ and $\nu_0(r)$ are the metric functions of the nonrotating solution, and $h_l(r)$, $m_l(r)$, $k_l(r)$ are the monopole ($l=0$) and quadrupole ($l=2$) contributions of second order in rotation respectively. The choice $k_0(r)=0$ is part of Hartle's choice of gauge.
 The Hartle metric \eqref{kingmet} is equivalent to second order to general metric \eqref{axisymstat} with the identifications (see eg \cite{Chandrasekhar:1974})
\begin{subequations}
\begin{align}
&e^\nu =  e^{\nu_0(r)} \big[ 1 + h_0(r) + h_2(r) \,P_2(\cos \theta)\big]\\
&e^\psi =  r \sin\theta\,  \big[1  + k_2(r)\, P_2(\cos \theta)\big]\\
&e^\alpha =  \sqrt{\frac{r}{r-2 m(r)}} \left\{1 +  \frac{m_0(r)+m_2(r)P_2(\cos \theta)}{r- 2 m(r)}\right\} \\
&e^\beta =  r \big[1  + k_2(r) \, P_2(\cos \theta)\big]\\
&\omega=\omega(r).
\end{align}
\end{subequations}
\subsection{Hartle's energy-momentum tensor}
Originally, Hartle's metric was paired with a perfect fluid energy-momentum tensor. We describe its construction here for completeness, but since we are interested in a system with anisotropic pressures we use a different method to construct and examine the energy-momentum tensor which is described in the following section.  With a background metric of the form
\begin{align}
ds^2 = - e^{2\nu_0(r)} \, dt^2 + \frac{dr^2}{1-\frac{2m(r)}{r}} + r^2 \, d\theta^2 + r^2 \, \sin^2\theta \, d\phi^2 .
\label{sphmetintro}
\end{align} and the unperturbed energy-momentum tensor $T^\mu_{~\nu}=diag(-\rho,p,p,p)$, Einstein's equations give the following  relationships:
\begin{align}
 \frac{\partial m}{\partial r} & = 4\pi r^2 \rho ,
\label{eq:Einstein-rhointro}
\\
\frac{\partial\nu_0}{\partial r} & = \frac{(m+4\pi r^3 p)}{r^2 \left(1-\frac{2m}{r}\right)} ,
\label{eq:Einstein-printro}
\\
- \frac{\partial p}{\partial r}&=\frac{ \left( m+4\pi r^3 p \right) \left( \rho+p \right)}{r^2 \left( 1 - \frac{2  m}{r} \right) }.
   \label{Eq:Forceintro}
\end{align}
Given an equation of state and appropriate boundary conditions, one may in theory solve this system for the unperturbed metric functions. In the notation of Hartle~\cite{Hartle:1967he}, the perturbed energy-momentum tensor is
\begin{align}
T^{\mu\nu} = (\EE + \PP) u^\mu u^\nu + \PP g^{\mu\nu},
\label{Tfluid}
\end{align}
where $\EE$ and $\PP$ are the energy density and pressure in the comoving frame of the rotating fluid, and $u^\mu$ is its four-velocity 
\begin{equation}
    u^t=\frac{1}{\sqrt{-g_{tt}-2\Omega g_{t\phi}-\Omega^2 g_{\phi \phi}}},\qquad u^\phi=\Omega u^t,\qquad u^r=u^\theta=0.
\end{equation}
To order $\Omega^2$,
\begin{align}
& \EE = \rho(r) + \EE_0(r) + \EE_2(r) P_2(\cos\theta),
\\
& \PP = p(r) + \PP_0(r) + \PP_2(r) P_2(\cos\theta) ,
\end{align}
where  $\EE_0(r)$, $\EE_2(r)$, $\PP_0(r)$, $\PP_2(r)$ are monopole and quadrupole perturbation functions of order $\Omega^2$, where in the case of these perfect fluid systems the rotation parameter $\Omega$ has a simple interpretation as a uniform angular velocity ($\Omega$ loses such a simple interpretation for vacuum energy type solutions, such as the pure vacuum Hartle-Thorne solution \cite{Hartle:1968si} and de Sitter like solutions\cite{UchiYoshida:2014,Uchikata:2015,Pani:2015,UchiPani:2016,2022PhRvD.105b4002B} because the four velocity drops out the energy-momentum tensor \ref{Tfluid}) . 
Note that in \cite{Hartle:1968si}, they define fractional changes
\begin{align}
& \PP = p(r) + (\rho+p)(\delta p_0(r) + \delta p_2(r) P_2(\cos\theta)),\\
& \EE = \rho(r) + \frac{d\rho}{dp}(\rho+p)(\delta p_0(r) + \delta p_2(r) P_2(\cos\theta)),
\end{align}
which are commonly used in other works. With Eq.~\eqref{Tfluid}, the Einstein tensor for Eq.~\eqref{kingmet}, appropriate boundary conditions, and the equation of state, one may solve for the perturbation functions.

\section{koosh}
The stringy equation of state $\Lambda_0=\Lambda_1$, $\Lambda_2=\Lambda_3=0$ is radically different than a perfect fluid equation of state $\Lambda_3=\Lambda_2=\Lambda_1=f(\Lambda_0)$. However, we find that if we examine the Einstein tensor order by order we can identify Hartle perturbation metrics which correspond to rotating kooshes and satisfy the stringy equation of state.
Keeping terms which are zeroth or first order in rotation, we find the metric is specified by
\begin{equation}
     ds^2=-\kappa^2  dt^2+\frac{1}{\kappa^2}dr^2+ r^2 d\theta^2+r^2 \sin^2\theta  d\phi^2-2 r^2\sin^2\theta\omega(r)d\phi dt.
   \label{kooshW1met}
 \end{equation}
 and the energy-momentum tensor has nonzero components
 \begin{align}
     T^t_{~t}=T^r_{~r}=-\frac{\lambda}{4\pi r^2},\\
     T^t_{~\phi}=-\frac{r \sin ^2(\theta ) \left(4 \omega '(r)+r \omega ''(r)\right)}{16 \pi }\\
     T^\phi_{~t}=-\frac{ \left(2 (2 \lambda -1) r \left(4 \omega '(r)+r \omega
   ''(r)\right)+8 \lambda  \omega (r)\right)}{32 \pi  r^2}
 \end{align}
 This suggests two special\footnote{As we will see in the next subsection, preservation of the equation of state is not enough to fully specify the frame dragging. However, because these examples for frame dragging lead to terms going to zero, the  other equations simplify and exact solutions for the other functions can be found.} frame dragging solutions: 
 \begin{align}
    T^t_{~\phi}=0 \rightarrow4 \omega '(r)+r \omega ''(r)=0\rightarrow \omega(r)=W_1+\frac{W_2}{r^3}\label{vacuumdrag1}    \\
    T^\phi_{~t}=0\rightarrow 
    \left(4 \omega '(r)+r \omega
   ''(r)\right)=\frac{-4 \lambda  \omega (r)}{(2 \lambda -1) r}\rightarrow  \omega(r)=\frac{W_a r^{-\sqrt{\frac{9-2 \lambda }{4-8 \lambda }}}+W_b
   r^{\sqrt{\frac{9-2 \lambda }{4-8 \lambda }}}}{r^{3/2}}.\label{alternatedrag1}
 \end{align}
 Notice that Eq. \eqref{vacuumdrag1} gives the same dragging as the vacuum Hartle-Thorne \cite{Hartle:1968si} and vacuum energy de-Sitter type solutions \cite{UchiYoshida:2014,Uchikata:2015,Pani:2015,UchiPani:2016,2022PhRvD.105b4002B}; we will therefore call it ``vacuum dragging." Interestingly, there is no Komar angular momentum Eq.~\eqref{KomarAngmom} associated with this frame dragging for the volume term of the Koosh as we have $ T^t_{~\phi}=0$. Note that in the case of the Hartle-Thorne solution and the vacuum energy de-Sitter type solutions the $W_2$ term is associated with an angular momentum concentrated inside the region of interest (such as a rotating star in the standard Hartle-Thorne picture \cite{Hartle:1968si} or a delta function in \cite{2022PhRvD.105b4002B}). The $W_1$ term can be associated with angular momentum concentrated outside the region of interest, specifically arising from a rotating eternal shell for the vacuum energy de-Sitter type solutions considered in \cite{UchiYoshida:2014,Uchikata:2015,Pani:2015,UchiPani:2016,2022PhRvD.105b4002B}.
\subsection{Preservation of the equations of state}
The full second order energy-momentum tensor is
\begin{align}
    T^t_{~t}=&\frac{-\lambda}{4\pi r^2}+\Bigg[\frac{r \left(2 \omega  \left(r \omega ''+4 \omega '\right)+r \left(\omega
   '\right)^2\right)}{48 \pi }-\frac{ m_0'}{4 \pi  r^2}+P_2(\cos\theta)\Bigg(\frac{\kappa ^2 
   \left(r k_2''+3 k_2'\right)}{4 \pi  r}-\frac{ 2k_2+ m_2'}{4 \pi  r^2}-\frac{3 m_2 }{4 \pi  \kappa ^2
   r^3}\nonumber\\&-\frac{r \left(2 \omega  \left(r \omega ''+4 \omega '\right)+r \left(\omega
   '\right)^2\right)}{48 \pi }\Bigg)\Bigg],\\
   T^t_{~\phi}=&-\{\frac{r \sin ^2(\theta ) \left(r \omega ''+4 \omega '\right)}{16 \pi }\},\\
   T^\phi_{~t}=&-\{\frac{\left(4 \lambda  \omega -\kappa^2 r \left(r \omega ''+4
   \omega '\right)\right)}{16 \pi  r^2}\},\\
   T^\phi_{~\phi}=&\Bigg[\frac{\kappa ^2 r^2 \left(r h_0''+h_0'\right)-r m_0'+m_0}{8 \pi 
   r^3}+\frac{h_2 \kappa ^2 r+m_2}{8 \pi  \kappa ^2 r^3}-\frac{r \left(2
   \omega  \left(r \omega ''+4 \omega '\right)+3 r \left(\omega
   '\right)^2\right)}{48 \pi }+\frac{P_2(\cos\theta)}{8 \pi   r^3} \Bigg(\nonumber\\& r \left( \kappa ^2 r \left(r
   \left(h_2''+k_2''\right)+h_2'+2 k_2'\right)-4 h_2- m_2'+\frac{r^3 \omega '}{6}
   \left(3 r \omega '+8 \omega \right)+ \frac{r^4 \omega  \omega ''}{3}\right)+
   m_2-\frac{4 m_2}{\kappa^2}\Bigg)\Bigg],\\
   T^r_{~r}=&\frac{-\lambda}{4\pi r^2}+\Bigg[\frac{ \kappa ^2 r^2 h_0'- m_0}{4 \pi  r^3}+\frac{r^2 \left(\omega '\right)^2}{48 \pi }-\frac{P_2(\cos\theta)}{4\pi r^3}\Bigg(3 h_2 r+2 k_2 r+ m_2+\frac{r^5
   \left(\omega '\right)^2}{12}- \kappa ^2 r^2 (h_2'+k_2')\Bigg)\Bigg],\\
   T^\theta_{~\theta}=&\Bigg[\frac{ \kappa ^2 r^2 (h_0'+  r h_0'')-  r m_0'+  m_0}{8
   \pi   r^3}-\frac{ h_2 \kappa ^2 r+ m_2}{8
   \pi  \kappa ^2 r^3}-\frac{ r^2 \left(\omega '\right)^2}{48
   \pi  }+\frac{P_2(\cos\theta)}{8 \pi   r^3}\Bigg(\nonumber\\& r \left( \kappa ^2 r^2 h_2''+ \kappa ^2 r h_2'+ \kappa ^2 r^2
   k_2''+2 \kappa ^2 r k_2'- m_2'+\frac{r^4 \left(\omega '\right)^2}{6}\right)-2 h_2
    r+ \left(1-\frac{2}{\kappa^2}\right) m_2\Bigg)\Bigg],\\
   T^r_{~\theta}=&r^2\kappa^2 T^\theta_{~r}=\Bigg[\frac{3  \sin (2 \theta ) \left(\kappa ^2 r^2
   \left(h_2'+k_2'\right)-r h_2 \kappa ^2 -m_2\right)}{16 \pi  r^2}\Bigg].
\end{align}

Here terms in big square brackets are second order, terms in curly brackets are first order, and we use the shorthand $\kappa$ from earlier.
 For a stationary axisymmetric metric of the form Eq.~\eqref{axisymstat} the eigenvalues of the energy-momentum tensor follow a pattern due to its block diagonal structure, and can be written as 
\begin{align}
    \Lambda_0=\frac{1}{2}\Big(T^t_{~t}+T^\phi_{~\phi}-\sqrt{(T^t_{~t}-T^\phi_{~\phi})^2-4T^\phi_{~t}T^t_{~\phi}}\Big)\\
    \Lambda_3=\frac{1}{2}\Big(T^t_{~t}+T^\phi_{~\phi}+\sqrt{(T^t_{~t}-T^\phi_{~\phi})^2-4T^\phi_{~t}T^t_{~\phi}}\Big)\\
    \Lambda_1=\frac{1}{2}\Big(T^r_{~r}+T^\theta_{~\theta}-\sqrt{(T^r_{~r}-T^\theta_{~\theta})^2-4T^\theta_{~r}T^r_{~\theta}}\Big)\\
    \Lambda_2=\frac{1}{2}\Big(T^r_{~r}+T^\theta_{~\theta}+\sqrt{(T^r_{~r}-T^\theta_{~\theta})^2-4T^\theta_{~r}T^r_{~\theta}}\Big)
\end{align}
When expanded to second order, the eigenvalues of this energy-momentum tensor are

\begin{align}
    \Lambda_0=&T^t_t-\frac{r \sin^2(\theta)(4\omega'+r\omega'')(4 \lambda \omega-\kappa^2 r (r \omega ''+4\omega '))}{64\pi\lambda},\\
   \Lambda_1=&T^r_{~r},\\
   \Lambda_2=&T^\theta_{~\theta},\\
   \Lambda_3=&T^\phi_{~\phi}+\frac{r \sin^2(\theta)(4\omega'+r\omega'')(4 \lambda \omega-\kappa^2 r (r \omega ''+4\omega '))}{64\pi\lambda}.
\end{align}

To this second order, we have $\Lambda_1=T^r_{~r}$ and $\Lambda_2=T^\theta_{~\theta}$ because the effects from the $T^r_{~\theta}T^\theta_{~r}$ cross term are pushed to higher order. 

One expression for eigenvectors in $(t,r,\theta,\phi)$ coordinates, correct to second order, is 
\begin{align}
    x_0^\mu=\left(1,0,0,\omega -\frac{\kappa ^2 r \left(r \omega ''+4 \omega
   '\right)}{4 \lambda }\right),\\
   x_1^\mu=\left(0,1,\frac{3 \sin (2 \theta ) \left(m_2-\kappa ^2 r \left(r
   \left(h_2'+k_2'\right)-h_2\right)\right)}{4 \kappa ^2 \lambda 
   r^2},0\right),\\
   x_2^\mu=\left(0,-\frac{3 \sin (2 \theta ) \left(m_2-\kappa ^2 r(
   r\left(h_2'+k_2'\right)-h_2 )\right)}{4 \lambda
   },1,0\right),\\
   x_3^\mu=\left(-\frac{r^3 \sin ^2(\theta ) \left(r \omega ''+4 \omega '\right)}{4
   \lambda },0,0,1\right).
\end{align}
Note that the $x_0^\mu,x_3^\mu$ have the zeroth order term and a first order term while the $x_1^\mu,x_2^\mu$ have the zeroth order term and a second order term. Notice that in the vacuum dragging Eq.~\eqref{vacuumdrag1} case $x_3^\mu$ is purely along the $\phi$ direction and in the alternate frame dragging case $x_0^\mu$ is purely along the $t$ direction.

Now that we have expressions for the components and eigenvalues in the energy-momentum tensor for arbitrary perturbation functions, we can now use a form of the stringy equation of state \ref{stringeos}, being $\Lambda_2=0,\Lambda_3=0,\Lambda_0-\Lambda_1=0$, to derive differential equations from which the perturbation functions should follow. Notice that each of these equations will separate into a monopole term and a quadrupole term.

Since $\Lambda_2$ and $\Lambda_3$ are separately zero, if we take their difference we also obtain $\Lambda_2-\Lambda_3=0$, which leads to the algebraic condition allowing for the elimination of $h_2$
\begin{align}
   h_2= \frac{- m_2}{\kappa ^2  r}-\frac{ r^4 \left(4 (7 \lambda -4) \left(\omega
   '\right)^2-\kappa ^2 r^2 \left(\omega ''\right)^2-8 \kappa
   ^2 r \omega ' \omega ''\right)}{24  \lambda  }
   \label{h2rule}.
\end{align}

Next, the equation of state \eqref{stringeos} requires the condition $\Lambda_0-\Lambda_1-2\Lambda_2=0$, using the quadrupole part of this condition and Eq.~\eqref{h2rule} gives a differential equation 
\begin{align}
m_2=&-\frac{\kappa ^2 r^2}{72 \lambda } \Big(-4 \kappa ^2 r^4 \omega ' \left((31 \lambda -16) r
   \omega ^{(3)}-\kappa ^2
   r^2 \omega ^{(4)}+2 (65 \lambda -33) \omega
   ''\right)+\nonumber\\&\kappa ^4 r^5 \left(r^2 (\omega ^{(3)})^2+r
   \omega '' \left(26 \omega ^{(3)}+r \omega
   ^{(4)}\right)+83 \left(\omega ''\right)^2\right)-12
   \lambda  m_2''+16 (5 \lambda -2) (7 \lambda -4) r^3 \left(\omega
   '\right)^2\Big), \label{m2rule}
\end{align}
where $(3),(4)$ specify third and fourth order derivatives with $r$, which specifies $m_2$. Finally, using the quadrupole part of  $\Lambda_0-\Lambda_1=0$ gives a differential equation for $k_2$
\begin{align}
   12 \kappa ^4 \lambda  r^2 \left(r k_2''+2 k_2'\right) =&72 \lambda  m_2-\kappa ^2 r^5 \Big(-2 (\lambda  (56 \lambda -47)+8)
   \left(\omega '\right)^2+2 \kappa ^2 (7 \lambda -3) r^2 \left(\omega
   ''\right)^2-\nonumber\\&\kappa ^4 r^2 \omega^{(3)}\left(r \omega
   ''+4 \omega '\right)-4 (\lambda  (34 \lambda -31)+7) r \omega ' \omega
   ''\Big) \label{k2rule}
\end{align}

For the monopole functions, we can use the remaining monopole term from $\Lambda_0-\Lambda_1=0$ to obtain an expression for $h_0'$, being
\begin{align}
    h_0'=\frac{  m_0-  r m_0'}{ \kappa ^2   r^2}+\frac{ r^3 \left(r \omega ''+4
   \omega '\right)^2}{24  \lambda }
   \label{h0rule}
\end{align}
If we take the derivative of this we can replace both $h_0'$ and $h_0''$ in the other monopole equation [with the condition Eq.~\eqref{h2rule}, $\Lambda_2=0$ and $\Lambda_3=0$ have identical forms], we obtain a final differential equation for $m_0$
\begin{align}
    m_0''=-\frac{r^3}{24 \lambda } \left(8 (13 \lambda -6) \left(\omega '\right)^2-\kappa ^2 r^2
   \omega '' \left(2 r \omega^{(3)}+13 \omega ''\right)-8 \kappa ^2 r
   \omega ' \left(r \omega^{(3)}+8 \omega ''\right)\right).
   \label{m0rule}
\end{align}

Importantly, the stringy equation of state can now be satisfied (to second order) regardless of what $\omega$ is, provided the second order functions obey the above rules. In theory, a system with any given function for $\omega$ could be used with Eqs. (\ref{h2rule})-(\ref{m0rule}) to find second order functions such that the equation of state is preserved. 
Presumably $\omega$ could be determined by some ansatz about the form of the energy-momentum tensor beyond the equation of state, such as the uniform angular velocity assumption in the standard Hartle framework.
\section{example frame draggings}
Despite the fact that the preservation of the equation of state is not sufficient to specify the frame dragging, we have isolated two cases which have interesting behavior: the vacuum frame dragging case Eq.~\eqref{vacuumdrag1} and the alternate case Eq.~\eqref{alternatedrag1}. In both of these cases we find exact solutions to the differential equations for the perturbation functions.
\subsection{Vacuum frame dragging}

For our first example, consider that the frame dragging is of the ``vacuumlike"  form Eq.~\eqref{vacuumdrag1}. Based on the behavior of the vacuumlike frame dragging in other systems, we might expect this solution to apply when the Komar angular momentum density is concentrated inside (for $W_2$) or outside (for $W_1$) the region of spacetime in which the background metric \eqref{uniaxialmet} applies. 

In the vacuum frame dragging case, Eq.~\eqref{m0rule} simply becomes
\begin{align}
    m_0''=-\frac{3 W_2^2}{r^5}
\end{align}
which may be trivially integrated to obtain
\begin{align}
    m_0=-\frac{W_2^2}{4r^3}+z_1 r+z_2\label{m0vac}
\end{align}
where $z_n$ are integration constants for the monopole functions. With $m_0$, Eq.~\eqref{h0rule} likewise simplifies and may be trivially integrated, giving
\begin{align}
    h_0'=\frac{1}{\kappa^2}\Big(\frac{z_2}{r^2}-\frac{W_2^2}{r^5}\Big)\\
    h_0=\frac{1}{\kappa^2}\Big(-\frac{z_2}{r}+\frac{W_2^2}{4r^4}\Big)+z_3.
\end{align}
This fully specifies the monopole functions.
The quadrupole functions are slightly more complicated. Equation~\eqref{m2rule} leads to
\begin{align}
    m_2=\frac{1}{6} \kappa ^2 r^2 m_2''+\frac{2 \kappa ^2 (3 \lambda -1)
   W_2^2}{r^3},\\
   m_2=q_1 r^{\frac{1}{2} \left(1-S\right)}+q_2 r^{\frac{1}{2} \left(1+S\right)}+\frac{2 \kappa ^2 (3 \lambda -1) W_2^2}{(4 \lambda
   -1) r^3}\label{m2vac}
\end{align}
where we introduce the shorthand $S=\sqrt{\frac{24}{\kappa^2}+1}$ and the integration constants for the quadrupole functions $q_n$. We use Eqs.~\eqref{h2rule} to obtain the expression for $h_2$
   \begin{align}
       h_2=\frac{3W_2^2}{2r^4}-\frac{m_2}{\kappa^2 r},
   \end{align}
where $m_2$ takes the form from Eq.~\eqref{m2vac}. Finally, we obtain
\begin{align}
    12 \kappa ^4 \lambda  r^2 \left(2 k_2'+r k_2''\right)=\frac{18 \lambda  \left(2 \lambda  (8 \lambda -5)+\frac{1}{4 \lambda
   -1}\right) W_2^2}{r^3}-72 \lambda  \left(q_1
   r^{\frac{1-S}{2}}+q_2 r^{\frac{S+1}{2}}\right),\\
   k_2=\frac{q_1 r^{-\frac{S+1}{2}}}{\kappa ^2}+\frac{q_2
   r^{\frac{S-1}{2}}}{\kappa ^2}+\frac{q_3}{r}+q_4+\frac{(4 (3-8
   \lambda ) \lambda +1)W_2^2}{8 \left(1-4\lambda\right)\kappa^2 r^4}
\end{align}
The nonzero eigenvalue becomes
\begin{align}
    \Lambda_0=&\Lambda_1=\frac{-\lambda}{4\pi r^2}+\Bigg[-\frac{z_1}{4\pi r^2}+\nonumber\\&\frac{P_2(\cos \theta)}{2 \pi  r^2} \left(\frac{q_3 (-3+2 \lambda )}{2
   r}-q_4+\frac{\left(q_1 r^{-\frac{1}{2} (1+S)}+q_2
   r^{\frac{S-1}{2}}\right) \lambda }{\kappa ^2}+\frac{W_2^2 \lambda 
   (-1+4 (3-4 \lambda ) \lambda )}{4 r^4 \left(1-4\lambda\right)\kappa^2}\right)\Bigg]\label{vaceig}
\end{align}
Notice how $W_1,~z_2,~z_3$ do not show up in the energy-momentum tensor eigenvalue. The entire energy-momentum tensor simplifies considerably, becoming
\begin{align}
    T^{t}_{~t}=T^{r}_{~r}=\Lambda_0=\Lambda_1,\\
   r^2\kappa^2T^{\theta}_{~r}= T^{r}_{~\theta}=\Bigg[\frac{-3\sin(2\theta)(W_2^2\lambda+q_3 r^3\kappa^2)}{16\pi r^5}\Bigg],\\
   T^{\phi}_{~t}=\frac{-\lambda(W_2+r^3 W_1)}{4\pi r^5},
\end{align}
with all other components vanishing to this order.

The $S_G$ and $\mathcal{J}$ scalars are
\begin{align}
    S_G=&\Bigg[\frac{ \log (\kappa )}{\kappa } \left(\frac{z_2}{ r^2}-\frac{W_2^2}{
   r^5}+P_2(\cos\theta) \left(\frac{1}{2} q_1 r^{-\frac{3+S}{2} }
   (1+S)-\frac{1}{2} q_2 r^{\frac{S-3}{2}} (S-1)+\frac{2W_2^2
   \kappa ^2}{r^5 (1-4 \lambda )}\right)\right)\Bigg],\label{SG1}\\
   \mathcal{J}=&\Big\{\frac{-3 W_2 \kappa \sin^2\theta}{2r^2}\Big\}.\label{J1}
\end{align}
It is noteworthy that the surface gravity parameter $S_G$ of the unperturbed koosh is zero because the $e^{2\nu}$ goes to a constant, so its derivative is zero (the Komar mass of the unperturbed koosh is also zero, as for the unperturbed koosh $S_G,\omega$ and $\mathcal{J}$ being integrated in Eq.~\eqref{KomarMass} are all separately zero). For the system with vacuum frame dragging, the $S_G$ is second order in rotation and $\mathcal{J}$ is first order in rotation.

\subsubsection{Divergences of perturbation functions in vacuum frame dragging}
If the integration constants $W_2,~z_2,~q_1,~q_3$ are nonzero, then there will be a divergence of the metric perturbation functions as $r\rightarrow0$. There will likewise be divergences as $r\rightarrow\infty$ in the quadrupole sector if $q_2$ is nonzero. There is a $r\rightarrow\infty$ divergence in the function $m_0$ if the integration constant $z_1$ is present , but it is noteworthy that $m_0$ enters into the metric \eqref{kingmet} in the combination $m_0/(r-2m)=m_0/(\kappa^2r)$, which has no divergences associated with $z_1$. Note however that for a given system any of these terms might be present if the background metric \eqref{uniaxialmet} is only valid over a certain domain, as is the case with global monopoles (which differs at small $r$) or something that is cut off by a shell (which would differ at large $r$). The integration constants $W_1,z_3,q_4$ do not cause any divergences in the metric functions. Strictly speaking, there is a divergence in the nonzero eigenvalue of the energy-momentum tensor Eq.~\eqref{vaceig} associated with $q_4$ as $r\rightarrow0$, but it has the same $1/r^2$ divergence as the background $\lambda/4\pi r^2$ term, so the $q_4$ term can still be considered small with respect to the background term. A similar formally divergent but small compared to the background term in Eq.~\eqref{vaceig} comes from $z_1$. The $q_1,q_3,W_2$ terms in Eq.~\eqref{vaceig} all diverge faster than the background term as $r\rightarrow0$, and the $q_2$ term will diverge as $r\rightarrow\infty$.

Independently of the divergences at large or small radii, there are particular values of $\lambda$ in which there are divergences. It is not surprising, given the pathological nature of the background line element  \eqref{uniaxialmet} in the $\lambda\rightarrow1/2,\kappa^2\rightarrow0$ quasiblack hole limit, that there are a multitude of divergences in the perturbation functions, associated with $W_2,z_2,q_1,q_2$. However, divergences due to these terms also show up in the eigenvalue Eq.~\eqref{vaceig} and the surface gravity parameter Eq.~\eqref{SG1}, which are scalars, showing that divergences caused by these integration constants in the quasiblack hole limit are not simply a coordinate artifact. Elimination of $W_2$ means that the frame dragging inside the quasiblack hole would go to a constant, which based on the behavior in the de Sitter type solutions might indicate the angular momentum is localized to the shell.

A more surprising feature is the fact that for nonzero $W_2$, the $m_2$ function (and hence $h_2$) will diverge at all radii when $\lambda=1/4$, in which the background metric exhibits no special behavior. This divergence at $\lambda=1/4$ also manifests in the $W_2$ term of Eq.~\eqref{vaceig}, and in the surface gravity parameter Eq.~\eqref{SG1}, indicating it has coordinate independent significance. A summary of divergences in scalar quantities under different conditions and the associated integration constants is given in Table \ref{divtable1}.

\begin{table}[]
\begin{center}
\begin{tabular}{ |c|c|c|c|c| }
 ~~~&$r=0$& $r=\infty$ & $\lambda=1/2$ & $\lambda=1/4$\\
 \hline
 $\Lambda$&$W_2,q_1,q_3$;$z_1^*,q_4^*$ &$q_2$ &$W_2,q_1,q_2$ &$W_2$ \\
 \hline
 $S_G$& $W_2,z_2,q_1$& $q_2$&$W_2,z_2,q_1,q_2$& $W_2$\\
 \hline
 $\mathcal{J}$&$W_2$ & & & \\
 \hline
 $\omega$& $W_2$& & & \\
\end{tabular}
\caption{Table showing integration constants which cause divergences in scalar quantities given certain conditions. An asterisk indicates that, while a divergence is formally present, it diverges in the same manner as the background and remains subdominant; $z_1$ and $q_4$ are associated with divergences of this type. $W_1$ and $z_3$ do not lead to divergences in any of these quantities. \label{divtable1}}
\end{center}
\end{table}
\subsection{Other frame dragging}
Recall that there is a second special frame dragging with regard to the off diagonal first order in rotation components, being Eq.~\eqref{alternatedrag1}. To simplify the notation, we introduce the shorthand 
\begin{align}
    Z=\sqrt{\frac{9-2 \lambda }{1-2 \lambda }}
\end{align}
such that
\begin{align}
    \omega=\frac{W_a r^{-Z/2}+W_b
   r^{Z/2}}{r^{3/2}}.
\end{align}
With this in mind, one may solve the monopole equations~(\ref{m0rule},\ref{h0rule}) and obtain
\begin{align}
    m_0=\frac{1}{24} W_b^2 (Z-3) r^Z-\frac{1}{24} W_a^2 (Z+3) r^{-Z}+r
   z_a+z_b\label{m0alt}\\
   h_0= \frac{W_a^2 (Z+3) r^{-Z-1}}{6 \kappa ^2 (Z+1)}+\frac{W_b^2 (Z-3)
   r^{Z-1}}{6 \kappa ^2 (Z-1)}-\frac{8 \lambda  W_a W_b+6 \kappa
   ^2 z_b}{6 \kappa ^4 r}+z_c
\end{align}
The solutions to the quadrupole equations (\ref{m2rule},\ref{h2rule},\ref{k2rule}) are
\begin{align}
    m_2=&q_a r^{\frac{S+1}{2}}+q_b r^{\frac{1-S}{2}}+\frac{1}{6}
   W_a^2 r^{-Z} (-2 \lambda  (Z+1)+Z+9)+\frac{1}{6} W_b^2 r^Z (2
   \lambda  (Z-1)-Z+9),\label{m2alt}\\
   h_2=& \frac{\frac{1}{12} W_a^2 r^{-Z-1} (2 \lambda -2 \lambda 
   Z+Z-9)+\frac{1}{12} W_b^2 r^{Z-1} (2 \lambda +(2 \lambda -1)
   Z-9)- q_a r^{\frac{S-1}{2}}-q_b r^{\frac{-S-1}{2}
   }}{\kappa ^2},\\
   k_2= &\frac{q_a r^{\frac{S}{2}-\frac{1}{2}} (-2 \lambda +(2 \lambda -1)
   S+25)}{\kappa ^4 (S-1) S}-\frac{q_b r^{\frac{1}{2} (-S-1)} (2
   \lambda +(2 \lambda -1) S-25)}{\kappa ^4 S
   (S+1)}-\frac{q_c}{r}+q_d+\nonumber\\&\frac{W_a^2 r^{-Z-1} \left(16
   \lambda ^2 (Z+1)-10 \lambda  (9+7Z)-9( 5Z-9)\right)}{12 \kappa ^2 (2
   \lambda -9) (Z+1)}+\frac{W_b^2 r^{Z-1} \left(16 \lambda ^2
   (Z-1)+10\lambda  (9-7 Z)-9 (5 Z+9)\right)}{12 \kappa ^2 (2 \lambda -9)
   (Z-1)}\nonumber\\&-\frac{4 \lambda  W_a W_b}{3 \kappa ^4 r}-\frac{4
   \lambda  W_a W_b \log (r)}{3 \kappa ^4 r}
\end{align}
where $S=\sqrt{\frac{24}{\kappa^2}+1}$ as before. Notice that the $q_a$ and $q_b$ terms in Eq. (\ref{m2vac}) and the $q_1,~q_2$ terms in (\ref{m2alt}) are analogous, as are the $z_1,~z_2$ and $z_a,~z_b$ terms in Eqs. (\ref{m0vac}) and (\ref{m0alt}). This is because they are associated with the homogeneous, or $\omega=0$, equations for $m_0$ and $m_2$, which become
\begin{align}
\text{Eq.(\ref{m0rule})}\rightarrow m_0''=0\\
\text{Eq.(\ref{m2rule})}\rightarrow m_2=\frac{\kappa^2 r^2}{6}m_2''
\end{align}
Because the terms are associated with $q_a,~q_b$ and $z_a,z_b$ terms are associated with the homogeneous $m_n$ equations analogous terms can show up in the $m_n$ functions for all frame draggings.

The nonzero eigenvalue of the energy-momentum tensor becomes
\begin{align}
    &\Lambda_0=\Lambda_1=\frac{-\lambda}{4\pi r^2}+\Bigg[\frac{W_a W_b \lambda }{6 \pi  r^3
   \kappa ^2}-\frac{z_a}{4 \pi  r^2}+\frac{\lambda  \left(r^{-3+Z} W_b^2+r^{-3-Z}
   W_a^2\right)}{12 \pi  \kappa ^2}+P_2(\cos \theta)\Bigg( \frac{q_c (3-2 \lambda )}{4 \pi 
   r^3}-\frac{q_d}{2 \pi  r^2}\nonumber\\&+\frac{W_a W_b \lambda  (5-2 \lambda +(6-4 \lambda ) \log
   (r))}{6 \pi  r^3 \kappa ^4}+\frac{\lambda}{2\pi 
   \kappa ^2}\Big(q_a r^{(S-5)/2}+q_b r^{-(S+5)/2}\Big)-\nonumber\\&\frac{16\lambda(5-2\lambda)}{\kappa^2} \left(\frac{r^{-3-Z}
  W_a^2}{96\pi(1+Z)}-\frac{r^{-3+Z} W_b^2}{96\pi(Z-1)}\right)-\frac{16 Z \lambda(3+4\lambda(\lambda-5))}{9+4\lambda(\lambda-5)}
   \left(\frac{r^{-3-Z} W_a^2}{96\pi(1+Z)}+\frac{r^{-3+Z}
   W_b^2}{96\pi(Z-1)}\right)\Bigg)\Bigg]
   \label{altereig}
\end{align}

The energy-momentum tensor follows the structure
\begin{align}
    & T^{t}_{~t}=T^{r}_{~r}=\Lambda_0=\Lambda_1,\\
   & r^2\kappa^2T^{\theta}_{~r}= T^{r}_{~\theta}=\frac{\sin(2\theta)}{8\pi r^2}\Bigg[\frac{3 q_c \kappa ^2}{2}+\frac{\left(3 \left(S-25+(S-1) \left(S \kappa
   ^2-2 \lambda \right)\right)\right) \left(q_a
   r^{\frac{1+S}{2}}+q_b r^{\frac{1-S}{2}}\right)}{4 S \kappa
   ^2}+\nonumber\\&\frac{2 W_a W_b \lambda  \log (r)}{\kappa ^2}-\frac{3\lambda}{4}
    \left(r^{-Z} W_a^2+r^Z
   W_b^2\right)+\frac{Z}{4}  \left(\frac{2\lambda^2-13\lambda}{9-2 \lambda }\right) \left(r^{-Z} W_a^2-r^Z
   W_b^2\right)\Bigg],\\
   & T^{t}_{~\phi}=-\frac{\left(\lambda  \sin ^2(\theta )\right) \left(W_a
   r^{-\frac{3}{2}-\frac{Z}{2}}+W_b
   r^{-\frac{3}{2}+\frac{Z}{2}}\right)}{4 \pi  \kappa ^2},
\end{align}
with all other components vanishing to this order. Notice that in this case as well as in the vacuum dragging case, we had $T^\phi_{~\phi}=0$ and $T^t_{~t}=\Lambda_0$. This is because we either had one or the other of $T^\phi_{~t}$ or $T^t_{~\phi}$ as zero for the particular frame dragging function. For other frame dragging functions for which the product $T^\phi_{~t}T^t_{~\phi}\ne0$, we no longer require $T^t_{~t}=\Lambda_0$ and $T^\phi_{~\phi}=0$ to satisfy the general expressions.

The $S_G$ and $\mathcal{J}$ scalars become 
\begin{align}
    S_G=&\frac{\log(\kappa)}{\kappa}\Bigg[\frac{1}{6} r^{-2+Z} W_b^2 (Z-3)-\frac{1}{6} r^{-2-Z} W_a^2
   (3+Z)+\frac{z_b}{r^2}+\frac{4 \lambda  W_a W_b}{3 r^2
   \kappa ^2}+P_2(\cos\theta) \Big(\frac{1}{2} q_a r^{-\frac{3}{2}+\frac{S}{2}}
   (1-S)\nonumber\\&+\frac{1}{2} q_b r^{-\frac{3}{2}-\frac{S}{2}}
   (1+S)+\frac{1}{12} \left((Z-1) \kappa ^2-8\right) \left(r^{-2-Z}
   W_a^2 (-1-Z)+r^{-2+Z} W_b^2 (1+Z)\right)\Big)\Bigg]\label{altersg}\\
    \mathcal{J}=&\frac{1}{4} \kappa  \sin ^2(\theta ) \left(W_b (Z-3)
   r^{\frac{Z-1}{2}}-W_a (Z+3) r^{\frac{1}{2} (-Z-1)}\right)
\end{align}
\subsubsection{Divergences in the alternate frame dragging}
\begin{figure}
    \centering
    \includegraphics{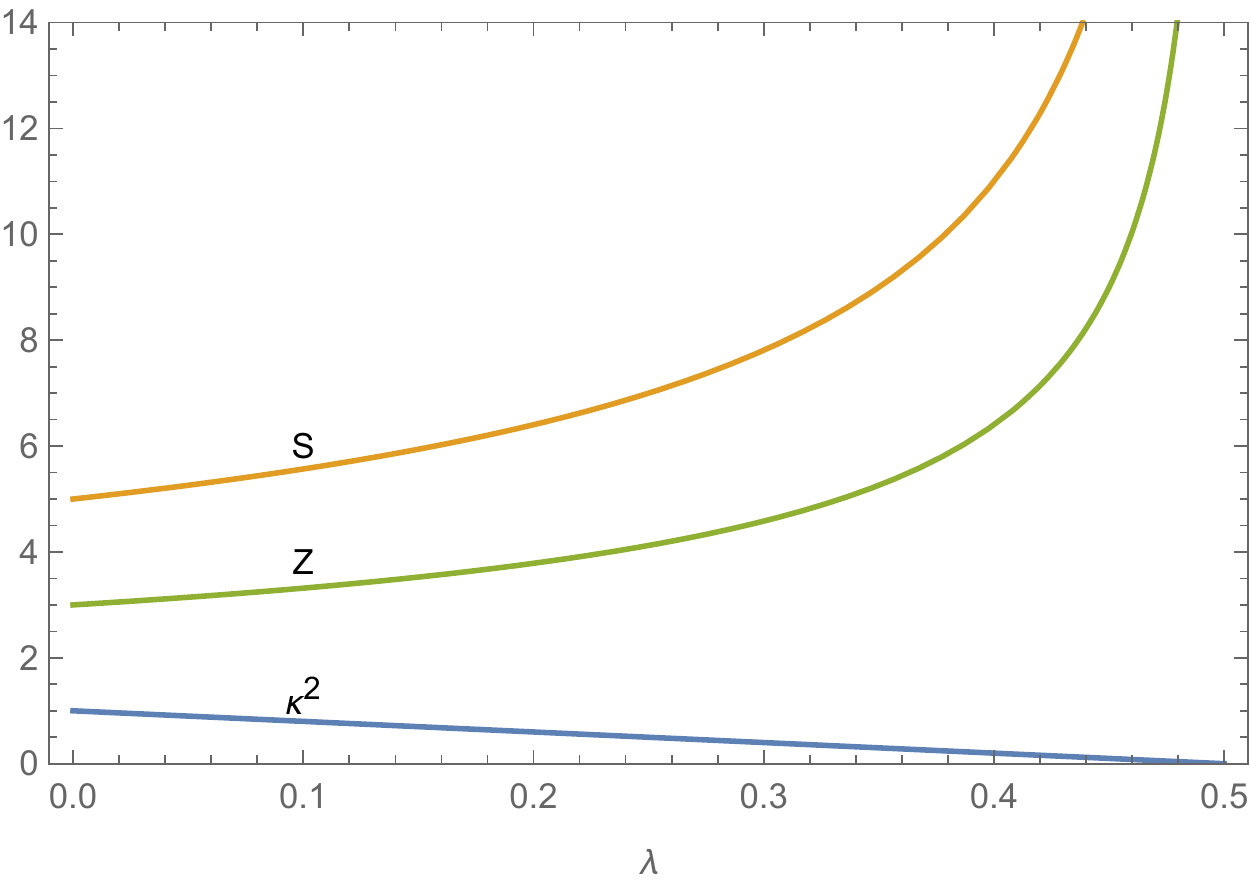}
    \caption{Graph showing the behavior of the shorthand items with respect to the parameter $\lambda$. Over the interval $0\le\lambda<1/2$ that we are interested in, $\kappa^2$ goes from 1 to 0, $Z$ goes from 3 to $\infty$, and $S$ goes from 5 to $\infty$}
    \label{params}
\end{figure}

Because the functions in the alternate frame dragging involve more notational shorthand to write in a reasonable amount of space, it is helpful to review how the different shorthand parameters are related to $\lambda$ before examining what divergences may exist. Plots of the shorthand paramaters are depicted in Fig. \ref{params}, where we see that $5\leq S\le\infty$ and $3\leq Z\le\infty$.
There are many terms in the perturbation functions and energy-momentum tensor components which have powers of $\kappa$ in denominators, which would of course lead to divergences in the quasiblackhole limit. Notice that within the eigenvalue Eq.~\eqref{altereig} and within the surface gravity parameter Eq.~\eqref{altersg}, both $W_a$ and $W_b$ have to be zero in order for these scalars to not diverge at all radii in the quasiblackhole limit, which implies no frame dragging could be present, or that the alternate frame dragging is not really compatible with a quasiblackhole configuration. The other terms in denominators i.e. $Z-1,Z+1,S,S-1,S+1,S^2-1,2\lambda-9$ do not have a zero over the $0\le\lambda<1/2$ interval, so there are no intermediate values of $\lambda$ causing divergences like $\lambda=1/4$ did for the vacuum dragging case.

There are however terms which diverge either as $r\rightarrow0$ or $r\rightarrow\infty$. Because $Z\ge3$, the presence of nonzero $W_a$ will cause divergences as $r\rightarrow0$ in the frame dragging, all second order functions, and all nonzero energy-momentum tensor components for general $\lambda$. The $W_b$ terms are typically associated with divergences as $r\rightarrow\infty$. However, for $\lambda=0,Z=3$, the $W_b$ term in frame dragging goes to a constant, and the $W_b$ terms in the monopole perturbation functions and energy-momentum tensor components/ eigenvalues vanish, but the quadrupole perturbation functions still have divergent $W_b$ terms as $r\rightarrow\infty$. Among the monopole integration constants, $z_a$ causes a divergence in $m_0$ as $r\rightarrow\infty$ [note the caveat that the combination in the line element $m_0/(r-2m)=m_0/(\kappa^2r)$ remains finite since $z_a$ and $z_1$ are analogous terms from the homogeneous solution] and a formally divergent but small compared to the background term in Eq.~\eqref{altereig}, $z_b$ causes a  divergence in $h_0$ as $r\rightarrow0$, and $z_c$ does not cause any divergences. Among the quadrupole constants, $q_a$ is associated with divergences as $r\rightarrow\infty$, $q_b$ and $q_c$ are associated with divergences as $r\rightarrow0$, and $q_d$ causes another formally divergent but small compared to the background term in the eigenvalue Eq.~\eqref{altereig}. We give a summary of divergences for the alternate frame dragging in Table \ref{divtable2}. It is important to reiterate that one may not care about divergent terms if they occur outside the domain where the background solution would be valid, such as $r\rightarrow \infty$ for objects cut off by a shell or $r \rightarrow 0$ for global monopole like objects with a smoothed core.
\begin{table}[]
\begin{center}
\begin{tabular}{ |c|c|c|c| }
 ~~~&$r=0$& $r=\infty$ & $\lambda=1/2$ \\
 \hline
 $\Lambda$&$W_a,W_a\times W_b,q_b,q_c$;$z_a^*,q_d^*$ &$W_b,q_a$ & $W_a,W_b,W_a\times W_b,q_a,q_b$  \\
 \hline
 $S_G$&$W_a,W_a\times W_b,z_b,q_b$ &$W_b,q_a$ &$W_a,W_b,W_a\times W_b,z_b,q_a,q_b$\\
 \hline
 $\mathcal{J}$&$W_a$ & $W_b$&\\
 \hline
 $\omega$& $W_a$& $W_b$ & \\
\end{tabular}
\caption{Integration constants which cause divergences in scalar quantities given certain conditions. An asterisk indicates that, while a divergence is formally present, it diverges in the same manner as the background and remains subdominant. Unlike the previous case, both frame dragging $W$ constants are associated with divergences, and we also have divergent terms proportional to the product $W_a\times W_b$. Additionally, there is no intermediate value of $\lambda$ associated with divergences in these quantities. Corresponding behavior with respect to conditions for divergences of the integration constants from the homogeneous sector of the equations ($z_a,z_b,q_a,q_b$ versus $z_1,z_2,q_1,q_2$) is evident in comparing the results with Table \ref{divtable1}. The integration constant $z_c$ has no associated divergences in these quantities. \label{divtable2}}
\end{center}
\end{table}

\section{Conclusion}
Within general relativity, rotating axisymmetric systems are of considerable interest. In this paper we show that a modified Hartle formalism (using the same form of perturbed metric but different conditions on the energy-momentum tensor) is capable of producing rotating solutions in a perturbative framework which preserve heavily anisotropic equations of state far different from the original application of perfect fluids. This is accomplished by deriving differential equations for the second order perturbation functions and presenting closed form solutions to a system which could be interpreted as describing a region of some global monopole or string cloud with rotation, although additional information beyond the equation of state is required to specify what frame dragging will be appropriate in a given physical situation. Examining specific physical situations and attempting to determine the appropriate frame dragging function (and by extension the other functions) is a possible avenue for future work. For instance, it seems likely that a system of the ``vaccuum" dragging case with $W_1$ only soldered to a Hartle-Thorne exterior may describe a stationary interior bounded by a rotating shell, although this would have to be verified. Additionally, one might consider a situation further akin to the original Hartle method and postulate a uniform angular velocity. This may be appropriate for a literal string cloud as it would prevent the strings from getting progressively more tangled, which could violate the assumption of a stationary system.  

One other possible extension of this work is examination of slowly rotating nonlinear electrodynamics monopoles. In the static case, Bardeen type nonsingular black holes can also arise from nonlinear electrodynamics theories (see e.g. \cite{AyonBeato:1998ub,AyonBeato:2000zs,Balart:2014cga} ) because of their [(11)(1,1)] Segre type, so rotating nonsingular black holes may also be amenable to this method. It is known that rotating versions of these black holes and nonlinear electrodynamics monopoles can be generated by the Newman-Janis algorithm (for instance \cite{2016EPJC...76..273A,Bambi:2013ufa}), but these rotating versions typically no longer follow the equation of state associated with the underlying nonlinear electrodynamics theory supposed to generate the static version \cite{Lombardo_2004,Beltracchi2021a}. Beyond the preservation of the equation of state, rotating systems generated by the Newman-Janis algorithm may contain singularities even if the original system was nonsingular \cite{Lamy:2018zvj}, which is an obvious drawback for modeling of nonsingular black holes. However, the koosh is also a Segre type [(11)(1,1)] system, can describe a monopole in a particular nonlinear electrodynamics theory \cite{2019arXiv191008166B}, and a modified version of the Hartle formalism was able to give rotating solutions which preserved its equation of state. Appropriate equations of state for nonsingular black holes or nonlinear electrodynamics monopoles may be derived from a Lagrangian, or may be phenomenologically derived from the nonrotating solution as in \cite{Beltracchi2021a}. That the modified Hartle method works for the koosh gives hope that rotating solutions for Bardeen type black holes or other nonlinear electrodynamics systems which satisfy the underlying equation of state may be found in a similar manner, at least within the slowly rotating nearly spherical limit .

\centerline{\bf Acknowledgement}
I would like to thank Paolo Gondolo and Emil Mottola for suggestions on this manuscript.
\bibliography{main.bib}

\end{document}